%
\let\useblackboard=\iftrue
%
%
\newfam\black

\input harvmac

\useblackboard
\message{If you do not have msbm (blackboard bold) fonts,}
\message{change the option at the top of the tex file.}
\font\blackboard=msbm10
\font\blackboards=msbm7
\font\blackboardss=msbm5
\textfont\black=\blackboard
\scriptfont\black=\blackboards
\scriptscriptfont\black=\blackboardss
\def\Bbb#1{{\fam\black\relax#1}}
\else
\def\Bbb{\bf}
\fi
\def\boxit#1{\vbox{\hrule\hbox{\vrule\kern8pt
\vbox{\hbox{\kern8pt}\hbox{\vbox{#1}}\hbox{\kern8pt}}
\kern8pt\vrule}\hrule}}
\def\mathboxit#1{\vbox{\hrule\hbox{\vrule\kern8pt\vbox{\kern8pt
\hbox{$\displaystyle #1$}\kern8pt}\kern8pt\vrule}\hrule}}

\def\subsubsec#1{\ifnum\lastpenalty>9000\else\bigbreak\fi
\noindent{\it{#1}}\par\nobreak\medskip\nobreak}
\def\yboxit#1#2{\vbotx{\hrule height #1 \hbox{\vrule width #1
\vbox{#2}\vrule width #1 }\hrule height #1 }}
\def\fillbox#1{\hbox to #1{\vbox to #1{\vfil}\hfil}}
\def\ybox{{\lower 1.3pt \yboxit{0.4pt}{\fillbox{8pt}}\hskip-0.2pt}}

\def\QZ{\Bbb{Z}}

\def\eps{\epsilon}


\def\CM{{\cal M}}

\def\CN{{\cal N}}

\

\def\drawbox#1#2{\hrule height#2pt 
        \hbox{\vrule width#2pt height#1pt \kern#1pt \vrule width#2pt}
              \hrule height#2pt}

\def\Asym#1#2{\vcenter{\vbox{\drawbox{#1}{#2}
              \kern-#2pt       
              \drawbox{#1}{#2}}}}

\def\p{\partial}

\def\frac#1#2{{#1 \over #2}}

\def\eps{\epsilon}



\def\cf{{\rm cf.}}  


\lref\IntriligatorPY{
  K.~Intriligator, N.~Seiberg and D.~Shih,
  ``Supersymmetry Breaking, R-Symmetry Breaking and Metastable Vacua,''
  JHEP {\bf 0707}, 017 (2007)
  [arXiv:hep-th/0703281].
}
\lref\IntriligatorCP{
  K.~Intriligator and N.~Seiberg,
  ``Lectures on Supersymmetry Breaking,''
  Class.\ Quant.\ Grav.\  {\bf 24}, S741 (2007)
  [arXiv:hep-ph/0702069].
}
\lref\IntriligatorDD{
  K.~Intriligator, N.~Seiberg and D.~Shih,
  ``Dynamical SUSY breaking in meta-stable vacua,''
  JHEP {\bf 0604}, 021 (2006)
  [arXiv:hep-th/0602239].
}
\lref\NelsonNF{
  A.~E.~Nelson and N.~Seiberg,
  ``R symmetry breaking versus supersymmetry breaking,''
  Nucl.\ Phys.\  B {\bf 416}, 46 (1994)
  [arXiv:hep-ph/9309299].
}

\lref\DineGM{
  M.~Dine, J.~L.~Feng and E.~Silverstein,
  ``Retrofitting O'Raifeartaigh models with dynamical scales,''
  Phys.\ Rev.\  D {\bf 74}, 095012 (2006)
  [arXiv:hep-th/0608159].
}
\lref\DineDZ{
  M.~Dine and J.~D.~Mason,
  ``Dynamical Supersymmetry Breaking and Low Energy Gauge Mediation,''
  arXiv:0712.1355 [hep-ph].
}
\lref\DineXT{
  M.~Dine and J.~Mason,
  ``Gauge mediation in metastable vacua,''
  Phys.\ Rev.\  D {\bf 77}, 016005 (2008)
  [arXiv:hep-ph/0611312].
}
\lref\AharonyDB{
  O.~Aharony, S.~Kachru and E.~Silverstein,
  ``Simple Stringy Dynamical SUSY Breaking,''
  Phys.\ Rev.\  D {\bf 76}, 126009 (2007)
  [arXiv:0708.0493 [hep-th]].
}

\lref\KaplunovskyRD{
  V.~S.~Kaplunovsky and J.~Louis,
  ``Model independent analysis of soft terms in effective supergravity and in
  string theory,''
  Phys.\ Lett.\  B {\bf 306}, 269 (1993)
  [arXiv:hep-th/9303040].
}

\lref\LouisUX{
  J.~Louis,
  ``Status Of Supersymmetry Breaking In String Theory,''
  talk given at the 1991 DPF conference, preprint SLAC-PUB-5645.
}

\lref\KachruEM{
  S.~Kachru, J.~McGreevy and P.~Svrcek,
  ``Bounds on masses of bulk fields in string compactifications,''
  JHEP {\bf 0604}, 023 (2006)
  [arXiv:hep-th/0601111].
}

\lref\GranaNY{
  M.~Grana, J.~Louis and D.~Waldram,
  ``Hitchin functionals in N = 2 supergravity,''
  JHEP {\bf 0601}, 008 (2006)
  [arXiv:hep-th/0505264].
}

\lref\GranaSN{
  M.~Grana, R.~Minasian, M.~Petrini and A.~Tomasiello,
  ``Generalized structures of N=1 vacua,''
  JHEP {\bf 0511}, 020 (2005)
  [arXiv:hep-th/0505212].
}

\lref\GranaSV{
  M.~Grana, R.~Minasian, M.~Petrini and A.~Tomasiello,
  ``Type II strings and generalized Calabi-Yau manifolds,''
  Comptes Rendus Physique {\bf 5}, 979 (2004)
  [arXiv:hep-th/0409176].
}

\lref\GranaBG{
  M.~Grana, R.~Minasian, M.~Petrini and A.~Tomasiello,
  ``Supersymmetric backgrounds from generalized Calabi-Yau manifolds,''
  JHEP {\bf 0408}, 046 (2004)
  [arXiv:hep-th/0406137].
}

\lref\GurrieriIW{
  S.~Gurrieri and A.~Micu,
  ``Type IIB theory on half-flat manifolds,''
  Class.\ Quant.\ Grav.\  {\bf 20}, 2181 (2003)
  [arXiv:hep-th/0212278].
}

\lref\SheltonCF{
  J.~Shelton, W.~Taylor and B.~Wecht,
  ``Nongeometric flux compactifications,''
  JHEP {\bf 0510}, 085 (2005)
  [arXiv:hep-th/0508133].
}

\lref\SheltonFD{
  J.~Shelton, W.~Taylor and B.~Wecht,
  ``Generalized flux vacua,''
  JHEP {\bf 0702}, 095 (2007)
  [arXiv:hep-th/0607015].
}

\lref\GranaHR{
  M.~Grana, J.~Louis and D.~Waldram,
  ``SU(3) x SU(3) compactification and mirror duals of magnetic fluxes,''
  JHEP {\bf 0704}, 101 (2007)
  [arXiv:hep-th/0612237].
}

\lref\LawrenceJB{
  A.~Lawrence, T.~Sander, M.~B.~Schulz and B.~Wecht,
  ``Torsion and Supersymmetry Breaking,''
  arXiv:0711.4787 [hep-th].
}

\lref\AchucarroSY{
  A.~Achucarro, S.~Hardeman and K.~Sousa,
  ``Consistent Decoupling of Heavy Scalars and Moduli in N=1 Supergravity,''
  arXiv:0806.4364 [hep-th].
}

\lref\deAlwisTG{
  S.~P.~de Alwis,
  ``On integrating out heavy fields in SUSY theories,''
  Phys.\ Lett.\  B {\bf 628}, 183 (2005)
  [arXiv:hep-th/0506267].
}
\lref\deAlwisTF{
  S.~P.~de Alwis,
  ``Effective potentials for light moduli,''
  Phys.\ Lett.\  B {\bf 626}, 223 (2005)
  [arXiv:hep-th/0506266].
}

\lref\BanksDH{
  T.~Banks,
  ``Remarks on M theoretic cosmology,''
  arXiv:hep-th/9906126.
}
\lref\BanksAY{
  T.~Banks,
  ``M-theory and cosmology,''
  arXiv:hep-th/9911067.
}
\lref\LawrenceZK{
  A.~Lawrence and J.~McGreevy,
  ``Local string models of soft supersymmetry breaking,''
  JHEP {\bf 0406}, 007 (2004)
  [arXiv:hep-th/0401034].
}
\lref\LawrenceKJ{
  A.~Lawrence and J.~McGreevy,
  ``Remarks on branes, fluxes, and soft SUSY breaking,''
  arXiv:hep-th/0401233.
}
\lref\VafaWI{
  C.~Vafa,
  ``Superstrings and topological strings at large N,''
  J.\ Math.\ Phys.\  {\bf 42}, 2798 (2001)
  [arXiv:hep-th/0008142].
}
\lref\CasasQI{
  J.~A.~Casas, Z.~Lalak, C.~Munoz and G.~G.~Ross,
  ``Hierarchical Supersymmetry Breaking and Dynamical Determination of
  Ccompactification Paameters by Nonperturbative Effects,''
  Nucl.\ Phys.\  B {\bf 347}, 243 (1990).
}
\lref\deCarlosDA{
  B.~de Carlos, J.~A.~Casas and C.~Munoz,
  ``Supersymmetry breaking and determination of the unification gauge coupling
  constant in string theories,''
  Nucl.\ Phys.\  B {\bf 399}, 623 (1993)
  [arXiv:hep-th/9204012].
}
\lref\BrignoleDJ{
  A.~Brignole, L.~E.~Ibanez and C.~Munoz,
  ``Towards a theory of soft terms for the supersymmetric Standard Model,''
  Nucl.\ Phys.\  B {\bf 422}, 125 (1994)
  [Erratum-ibid.\  B {\bf 436}, 747 (1995)]
  [arXiv:hep-ph/9308271].
}
\lref\BrignoleFB{
  A.~Brignole, L.~E.~Ibanez, C.~Munoz and C.~Scheich,
  ``Some Issues In Soft Susy Breaking Terms From Dilaton / Moduli Sectors,''
  Z.\ Phys.\  C {\bf 74}, 157 (1997)
  [arXiv:hep-ph/9508258].
}
\lref\LouisHT{
  J.~Louis and Y.~Nir,
  ``Some phenomenological implications of string loop effects,''
  Nucl.\ Phys.\  B {\bf 447}, 18 (1995)
  [arXiv:hep-ph/9411429].
}
\lref\GomezReinoWV{
  M.~Gomez-Reino and C.~A.~Scrucca,
  ``Constraints for the existence of flat and stable non-supersymmetric vacua
  in supergravity,''
  JHEP {\bf 0609}, 008 (2006)
  [arXiv:hep-th/0606273].
}
\lref\GomezReinoQI{
  M.~Gomez-Reino and C.~A.~Scrucca,
  ``Metastable supergravity vacua with F and D supersymmetry breaking,''
  JHEP {\bf 0708}, 091 (2007)
  [arXiv:0706.2785 [hep-th]].
}
\lref\SunNH{
  Z.~Sun,
  ``Continuous degeneracy of non-supersymmetric vacua,''
  arXiv:0807.4000 [hep-th].
}
\lref\CveticMH{
  M.~Cvetic and T.~Weigand,
  ``A string theoretic model of gauge mediated supersymmetry beaking,''
  arXiv:0807.3953 [hep-th].
}

\Title{\vbox{\baselineskip12pt
\hbox{BRX TH-599}}}
{\vbox{\centerline{F-term SUSY Breaking and Moduli}}}
\vskip -2em
\centerline{Albion Lawrence} 
\medskip
\centerline{{\it Theory Group, Martin Fisher School of
Physics, Brandeis University,}}
\centerline{{\it  MS057, 415 South St., Waltham, MA 02454 USA}}

\bigskip
\noindent
We discuss the coupling of heavy moduli fields to light fields when the dynamics of the latter, 
absent such couplings, yields metastable vacua.  We show that the 
survival of the vacuum structure of the local
model depends nontrivially on the cross-couplings of the two sectors.
In particular we find that for "local" models (such as those realized by D-branes in type
II string theories) with metastable vacua breaking supersymmetry via F-terms,
cross-coupling of the two sectors at an
intermediate scale can push the metastable
vacuum outside of the regime of the effective field theory.  We parametrize
the region in which the metastable vacua are safe.
We the show that sufficiently small cross-couplings can be made natural.
Finally, we briefly discuss the role of moduli in stringy realizations of "retrofitted"
SUSY-breaking sectors.

\medskip
\Date{\number\day\ August 2008}


\newsec{Introduction}

Supersymmetric field theories with dynamical or spontaneous SUSY-breaking
via F-terms generically have an effective field theory description as an O'Raifeartaigh-like model.
Such models consist of scalar fields which spontaneously break supersymmetry due to a 
"rank condition" on the superpotential.  The classic example is the following
superpotential for three chiral scalars:
\eqn\orsuper{
	W(X,A,B) = h X A^2 + m AB + \mu X 
}
If $y = h\mu/m^2 < 1$, this has a single SUSY-breaking vacuum at $A = B = 0$, $F_X = \mu$
(see for example \refs{\IntriligatorPY,\IntriligatorCP}\ for a thorough discussion of this model).
For any value of $y$, $X$ is a flat direction at tree level and is stabilized 
at the $U(1)_R$-preserving point $X = 0$
by the one-loop Coleman-Weinberg potential.
One may also deform this model by a small R-breaking term, 
$\delta W = \half\eps X^2$, which leads to a SUSY-preserving vacuum
at large $X = - \mu_0/\eps$ 
and leads to a metastable SUSY-breaking vacuum at $X \sim \eps f(h,m,\mu)$.

In string theory, the couplings $m,\mu,h$
generally depend on moduli such as parameters of the metric of the compactification
manifold. For example, in the local orientifold model of
\refs{\AharonyDB}, $X$ is an open string field, and 
the analog of $\mu$ is generated by D-brane instantons 
which depend on a K\"ahler class of the orientifold. This K\"ahler class will
become a dynamical modulus in a compact model.
We might expect that if the geometric moduli are stabilized at a high scale,
they (a) do not interfere significantly  with the SUSY-breaking dynamics, and 
(b) do not acquire significant F-terms. One could then engineer a 
scenario in which the SUSY breaking in the theory \orsuper\ is communicated to
a supersymmetric extension of the Standard Model via gravity or gauge or anomaly
mediation.

These conditions would seem to be a natural consequence of Wilsonian decoupling,
but one must take a little care.  To see this, consider a quadratic theory with
two scalar fields and the following mass matrix:
\eqn\massarray{
	\CM^2 = \left( \matrix{ M^2 & \gamma \cr \gamma & m_{light}^2 }
	\right)
}
Here $M \gg m_{light}$, so it appears that we have a heavy "modulus" and a light field,
and we can integrate the modulus out by setting it to zero.   However,
if the "cross-coupling" term $\gamma$ is larger than the {\it intermediate}\ scale
$\sqrt{Mm_{light}}$, the system is tachyonic. We will discuss further examples
in \S2. Note that in this model, the
tachyon can be removed by imposing a $\QZ_2$ symmetry which forces $\gamma = 0$.

In \S3\ we will discuss the case of light fields in globally supersymmetric models
whose dynamics, when decoupled from
moduli, yield metastable vacua which break SUSY through F-terms.  We will show
that when the simplest class of models is coupled to moduli, there is a lower 
bound on the cross-coupling beyond which
the vacuum is pushed out of the range of effective field theory.\foot{There have been
a number of general discussions of when SUSY-breaking vacua exist in local and globally
supersymmetric models -- see for example \refs{\GomezReinoWV,\GomezReinoQI}.
There have also been some general discussions of subtleties of integrating out
heavy fields in local and global SUSY models in 
\refs{\deAlwisTF\deAlwisTG-\AchucarroSY}. We are concerned with the particular issue 
of the effects of coupling light fields with apparent SUSY-breaking dynamics to heavy moduli.}
In \S4\ we will discuss natural values of the cross-coupling terms and show that for
"geometric" moduli which have large ranges, the cross-couplings can
be naturally suppressed.

We will conclude by reviewing another possibility for generating a
sector such as \orsuper\  \refs{\DineGM\DineDZ-\DineXT}, 
in which the dimensionful couplings are generated by
nonperturbative gauge dynamics.  In string theory, $X$ will 
typically be a geometric modulus itself; this modulus 
is stabilized by a one-loop mass, and acquires an F-term.  Such models can be
realizations of "moduli domination" \refs{\KaplunovskyRD\CasasQI\deCarlosDA\BrignoleDJ-\BrignoleFB}.  

\newsec{Nonsupersymmetric examples}

Cross-couplings between "light" and "heavy" fields can change the vacuum structure of
a theory if those couplings are set by an intermediate mass scale.
Let us consider examples with two real scalar fields, a "light" field
$x$ and a heavy "modulus" $\phi$. We will consider quadratic kinetic terms,
and will assume that we have included all quantum corrections in our
effective action (although we will, for computational ease, write 
fairly nongeneric potentials).

First, consider the potential
\eqn\quartic{
	V = \frac{1}{4} \lambda x^4 + \half m^2 x^2 +  \gamma x \phi + \half M^2 \phi^2
}
where $M \gg m$.  If we integrate out $\phi$, we find the modified potential
\eqn\quartictwo{
	V = \lambda x^4 + \half (m^2 - \frac{\gamma^2}{M^2}) x^2\ .
}
Assume $m^2 > 0$, so that if we naively set $\phi = 0$, the potential \quartic\
would have a single vacuum at the origin $x = 0$.  If $\gamma$ is larger than the intermediate
scale $mM$, the actual vacuum structure will change, with $x = 0$ becoming
an unstable local maximum of the potential.

A more generic example begins with the following relatively generic potential for the light field $x$:
\eqn\generalpot{
	V = \frac{1}{4} \lambda x^4 + \frac{1}{3} g x^3 + \half m x^2
}
where $\lambda > 0$. This has extrema at 
\eqn\genpotext{
	x = 0, x_- = - \frac{g}{2} \pm \sqrt{\frac{g^2}{4 \lambda^2} - \frac{m^2}{\lambda}}
}
So long as $m^2 \leq g^2/(4\lambda)$, there are three extrema, two local minima 
and a local maximum. For example, if $m^2 > 0$, the minima are at $x_-$ and $0$, with $x_+$
the local maximum: the metastable local minimum at $x_-$ coalesces with the maximum at $x = 0$
and disappears as we dial $m^2 > g^2/4\lambda$.

Now, let us couple \genpotext\ to a "heavy" modulus $\phi$.  We will assume that the 
marginal and relevant couplings between $\phi$ and $x$ are linear in $\phi$ (this
is not generic -- we are making this assumption for ease of illustration, though
one might make the higher-order terms in $\phi$ small using the considerations of \S4):
\eqn\coupledgen{
	V = \frac{1}{4}\lambda x^4 + \frac{1}{3} \left( g + g_1 \phi\right) x^3
		+ \half (m^2 + m_1 \phi) x^2 + \gamma \phi x + \half M^2 \phi^2
}
In this case, we can integrate out $\phi$ exactly to find that:
\eqn\newcoupledgen{
	V = - \frac{g_1^2}{18 M^2} x^6 - \frac{g_1 m_1}{6 M^2} x^5
		+ \frac{1}{4} \lambda_{eff} x^4 + \frac{1}{3} g_{eff} x^3 + 
		\frac{1}{2} m_{eff}^2 x^2
}
where
\eqn\effcouple{
\eqalign{
	\lambda_{eff} & = \lambda - \frac{4 g_1\gamma}{3 M^2} - 
		\frac{m_1^2}{M^2}\cr
	g_{eff} & = g - \frac{3 m_1 \gamma}{2 M^2} \cr
	m_{eff}^2 & = m^2 - \frac{\gamma^2}{M^2}
}}
The $x^6$ and $x^5$ terms will generally be important compared to the
lower-order terms only when $x$ is of order $M$ or the initial couplings $\lambda, g, m$
are extremely small.  

Concentrating on the quartic and lower effective interactions, it becomes clear that while one
may change the sign of the quadratic term by choosing $\gamma$ at the intermediate
scale $m M$, even then it will be difficult to change the sign of the cubic and quartic terms,
or to achieve $m_{eff}^2 \geq g_{eff}^2/(4\lambda_{eff})$.  For these latter changes,
the massive coupling $m_1$ must be of order $Mg/m$ 
or the dimensionless coupling $g_1$ is extremely large, of
of order $\lambda M/m$).  Nonetheless, the sign change of the quadratic term will
have the effect of rendering the $x = 0$ vacuum unstable and the vacua at $x_{\pm}$
stable.  Inspired this, we will now discuss supersymmetric vacua with metastable 
SUSY-breaking states, and ask when the coupling to moduli can change the vacuum structure.

\newsec{Supersymmetric examples}

In this section we will consider globally supersymmetric
$N=1$ theories which break supersymmetry spontaneously through
F-terms before coupling to moduli.  These models can be taken to
represent the dynamics of open strings on D-branes which are
localized in the internal geometry of a type II compactification; of course,
there could be other representations.
Before coupling to moduli,  the SUSY-breaking vacua of these models may be completely
stable or metastable; generically, the latter occurs when one perturbs
the theory by operators, such as the superpotential term $\eps X^2$ added to \orsuper, which
break the $U(1)_R$ symmetry of the theory \refs{\IntriligatorPY,\IntriligatorCP,\IntriligatorDD}.

We will focus on a particular form of the cross-coupling (arising from letting the
SUSY-breaking coupling in the superpotential depend on a heavy modulus), and 
find that this coupling can change the vacuum structure. 
First, either the cross-coupling or
the moduli mass term will break the $U(1)_R$ symmetry of the SUSY-breaking sector, 
so that even if the model \orsuper\ has a SUSY-breaking global minimum, the full theory
will have supersymmetric minima.  Furthermore, the cross-couplings
can push the SUSY-breaking metastable state out of the range of the effective field theory.  
We will discuss this in detail in the context of a K\"ahler-stabilized Polonyi model, and find that
the physics of the SUSY-breaking sector can be significantly changed when the moduli mass $M$
is much smaller than the UV mass scale $\Lambda$ parametrizing the suppression of
higher-order terms in the K\"ahler potential.\foot{The recent work \refs{\CveticMH}\ includes
the effects of heavy moduli in generating a stable potential for the light fields.  This
work does not include the cross-couplings discussed in this section; however it is
quite possible that they can be suppressed using the considerations of \S4.}  
We will briefly discuss the explicit
case of the O'Raifeartaigh model as well (which at sufficiently low energies
and for a range of parameters, is well-described by a K\"ahler-stabilized Polonyi model).\foot{Recent
results on the stability of SUSY-breaking vacua \refs{\SunNH}\ concentrate on tree-level
physics with a minimal K\"ahler potential, so that this falls outside of the considerations of that work.}

\subsec{Polonyi model}

We will consider a theory with K\"ahler potential
\eqn\kskp{
	K = |X|^2 + |Z|^2 - \frac{c}{\Lambda^2} |X|^4
}
where $\Lambda$ is a cutoff scale such as the Planck scale, and $c$ is a dimensionless
number,
and superpotential:
\eqn\Polonyisuper{
	W = (\mu_0 + \mu_1 Z) X - \half \epsilon X^2 + \half M Z^2
}
This is mean to model the dependence of $\mu X$ on moduli.

If we decouple the two fields by setting $\mu_1 = 0$, $X$ has a SUSY vacuum at $X = \mu_0/\eps$.
So long as
\eqn\nocouplingcond{
	\eps < \frac{\sqrt{2c}\mu_0}{\Lambda}\ ,
}
then there is a metastable SUSY-breaking vacuum at
$X = \frac{\bar{\eps}\Lambda^2}{2\mu_0 c}$, and
\eqn\decoupledf{
	F_x^* = \mu_0 - \frac{\eps^2 \Lambda^2}{2c\mu_0}
}
The second term remains smaller than the first if \nocouplingcond\ is satisfied.
For larger values of $\eps$, the vacuum would be at 
$X > \frac{\Lambda}{\sqrt{2c}}$: the kinetic term for $X$ that arises from
\kskp\ alone flips sign, and higher-order terms in the K\"ahler potential must be taken
into account. This region is outside of the domain of our effective theory.

If we include the coupling to $Z$, we can integrate out $Z$ as follows \refs{\deAlwisTG}:\foot{
In this case, solving $\p_Z W = 0$ for $Z$ yields the same answer.  In general, as pointed
out in \refs{\deAlwisTG}, this does not give the full potential, though the difference will
be suppressed by powers of $M$.} The equations of motion for the components of $Z$ are:
\eqn\correctintegout{
\eqalign{
	F_z^* & = \p_z W = \mu_1 x + M z \cr
	- \p^2 z & = \mu_1 F_x + M F_z
}}
If we assume that the fields are constant, we find that the potential is:
\eqn\lowenergypot{
	U(x) = \frac{\biggl| \mu_0 - \left(\frac{\mu_1^2}{M} + \eps\right) x \biggr|^2}{1 + \frac{|\mu_1|^2}{M^2}
		 - \frac{2c}{\Lambda^2}|x|^2}
}
where $x$ is the scalar component of $X$. We will assume that $c,\mu_0,\mu_1$ and
$\epsilon$ are all real.

The potential \lowenergypot\ is singular at: 
\eqn\poleloc{
	|x_{sing}| = \frac{\Lambda}{\sqrt{2c}}\left(1 + \frac{|\mu_1|^2}{M^2}\right)\ .
}
If all of the parameters are real, as we supposed above, the solutions
to $\p_x U = 0$ are real. There is a supersymmetry-preserving solution at:
\eqn\susyloc{
	x_{susy} = \frac{\mu_0}{\frac{\mu_1^2}{M} + \eps}\ ,
}
and SUSY-breaking solutions to $\p_X U = 0$ at $x = \infty$
and at
\eqn\susybrk{
	x_{sb} = \frac{\Lambda^2}{2 c \mu_0} \left(1 + \frac{\mu_1^2}{M^2}\right)
		\left(\eps + \frac{\mu_1^2}{M} \right)
}
The solution to $\p_x U = 0$ between the origin and
$x_{sing}$ will be stable. The dynamics at larger field values is out
of range of our effective theory; higher-order corrections in the 
K\"ahler potential and superpotential will become important. 

$X= x_{sb}$ will be the stable vacuum between the origin and $x_{sing}$ if: 
\eqn\survivalcond{
	\eps M + \mu_1^2 < \sqrt{2c} \frac{M}{\Lambda} \mu_0 \left(1 + \frac{\mu_1^2}{M^2}\right)^{1/2}
}
Let us assume that $\mu_0^{1/2}, \mu_1, \eps \ll M,\Lambda$.
If $\eps$ is such that the SUSY-breaking vacuum would be stable if $\mu_1 = 0$, there
is still a condition on $\mu_1$: if $M/\Lambda \ll 1$, $\mu_1$
must be considerably smaller than the mass scale set by $\mu_0$, lest the coupling
to $Z$ pushes the SUSY-breaking vacuum out of the regime of our effective theory.
In particular, since $\sqrt{2c} \mu_0/\Lambda \sim m_x$ is the
mass of $X$ before coupling to $Z$, this means that the SUSY-breaking vacuum is under control
if $\mu_1^2 < m_x M$.

We can also compute the corrections to the F-terms that arise from coupling $X$ and $Z$.  
Using \correctintegout\ and
\eqn\otherfterm{
	F_x^* = \mu_0 + \mu_1 Z - \eps X\ ,
}
we find after a few lines of algebra that:
\eqn\fterms{
\eqalign{
	F_x^* & =  \frac{\mu_0}{1 + \frac{\mu_1^2}{M^2}} - 
	\frac{\Lambda^2}{M^2}
		\left(\eps + \frac{\mu_1^2}{M}\right)^2 \frac{1}{2c\mu_0} \cr
	F_z^* & = - \frac{\mu_1}{M} F_x^*
}}
So long as \survivalcond\ is satisfied, we can show the second term in the expression
for $F_x^*$ is smaller than the first, and in particular the effect of coupling to $Z$ is small.
Furthermore, so long as $M$ is larger than the parameters $\mu_0^{1/2}, \mu_1$ of the local
model, $F_z$ remains much smaller than $F_x$. On the other hand, if
$M \sim \Lambda$, and the inequality \survivalcond\ is close to being saturated, the F-term for
$X$ is reduced substantially from the result $F_x^* = \mu_0$ of the local model.

We have focused on a particular form of the cross-coupling in globally supersymmetric
models.  It would be interesting to study cross-couplings in the K\"ahler potential itself.
It would also be interesting to study this question in the context of supergravity --
for example, even if the K\"ahler potential and superpotential factorize, supergravity
effects will typically induce cross-couplings; furthermore, in may interesting examples,
the modulus $Z$ will be stabilized by supergravity effects. For the theory above,
supergravity will not induce any quadratic cross-couplings between $X$ and $Z$,
and we expect the effects to be small if $\Lambda \ll m_{pl}$.

\subsec{The O'Raifeartaigh model}

Let us consider the O'Raifeartaigh model coupled to a modulus $z$:
\eqn\ormodulipot{
	W = h X A^2 + m AB - (\mu_0 + \mu_1 Z) X + \half M Z^2
}
with canonical kinetic terms for $X, Z$.
In principle we should let $m,h$ also depend on moduli; we avoid this for simplicity's sake.
At energies below $m$, we can integrate out $A,B$ and $Z$.  For $h\mu_0/m^2 \ll 1$,
when we are studying a SUSY-breaking vacuum close to $X = 0$,
the "Coleman-Weinberg" potential which arises from integrating out $A,B$ 
can be expressed as a correction to the K\"ahler potential. (See for example
\refs{\IntriligatorPY,\IntriligatorCP}\ for a thorough discussion of the Coleman-Weinberg potential
in this model.)

Since $Z$ only appears quadratically in the superpotential
it can be integrated out classically.  Let us first integrate out $A,B$ (which we can do first 
because these fields do not couple to $Z$.)  If $X$ remains close to the
origin, the result is essentially a Polonyi model of the type discussed above
(there is also an order $h^2$ shift in the coefficient of $X^2$ in \kskp);
the $|X|^4$ term will be of the form \kskp, with $2c = \frac{|h|^4}{64\pi^2}$ 
and $\Lambda^2 = m^2$. If $m, \mu_0^{1/2}, \mu_1 \ll M$, then the SUSY-breaking 
vacuum in the O'Raifeartaigh model is safely inside the realm of the original
effective field theory.

\newsec{Stability and naturalness}

When are conditions such as \survivalcond, or 
$\gamma < mM$ in \quartic, satisfied? One possibility arises if, as
with axions, the range of the moduli is large while the potential is generated
at a somewhat lower scale.  For example,  geometric moduli have a kinetic term of the form
\eqn\modkin{
	S = m_{pl,4}^2 \int d^4 x G_{ab}(\phi) \p \phi^a \p \phi^b
}
where $\phi$ are dimensionless, since they parametrize the metric of the compactification manifold.
In many models such as type II flux compactifications or heterotic M-theory compactifications
on a Calabi-Yau times a large interval, potentials then arise either from fluxes 
from D-brane instantons, or from gauge instantons, and have the form
\eqn\modpot{
	V = M^4 v(\phi)
}
where $M \ll m_{pl,4}$.
When we rescale $\phi^a \to z^a = m_{pl,4} \phi^a$, where $z^a$ are dimension-1 4d scalars with 
canonical kinetic terms, we find that
\eqn\modpotresc{
	V = M^4 v\left(\frac{z}{m_{pl,4}}\right)
}
Thus interactions are suppressed by powers of $1/m_{pl,4}$.\foot{See \refs{\KachruEM}\ for a discussion
of this phenomenon in type II flux compactifications, and \refs{\BanksDH,\BanksAY}\ for a similar
argument in the context of heterotic M-theory.} In many of these examples, this suppression is
the result of an $\CN=2$ supersymmetry (which would forbid moduli masses) that is
broken by fluxes at a scale lower than the Kaluza-Klein scale \refs{\VafaWI\LawrenceZK
\LawrenceKJ-\LawrenceJB}, or broken by D-branes and orientifolds
which are local on the internal manifold and so are expected to induce 
$\CN=1$ masses suppressed by volume factors.  Note that if we are interested
in the effects of Planck-suppressed couplings, we will also have to take into account
additional terms in the potential which arise from supergravity corrections.

Similarly, assume that supersymmetry-breaking dynamics arise from "local models"
of open string fields trapped on D-branes (or in the case of heterotic M-theory, from
the $E_8$ "walls").  Tree-level couplings to dimension-$\Delta$ operators $\CO$ in the
open string sector take the form:
\eqn\openclosed{
	\delta L = \frac{m_s^{\Delta - 4}}{g_s} f(\phi)\CO = \frac{m_s^{\Delta - 4}}{g_s} 
	f\left(\frac{z}{m_{pl,4}}\right)\CO\ ,
}
where $m_s$ is the string scale (or in heterotic M-theory, the 11d Planck scale)
which is generally less than the 4d Planck scale.
Instanton-generated couplings such as those in \refs{\AharonyDB}\ will have the form:
\eqn\openclosednp{
	\delta L = \frac{M_{np}^{\Delta - 4}}{g_s} f(\phi)\CO = \frac{M_{np}^{\Delta - 4}}{g_s} 
	f\left(\frac{z}{m_{pl,4}}\right)\CO\ .
}
where $M_{np}$ is nonperturbatively small compared to $m_s, m_{pl,4}$.

In particular, consider the case \openclosednp, and expand the theory
in small fluctuations about the appropriate minimum of \modpotresc.  
The SUSY-breaking parameter will be
$\mu_0 \sim M^2$, while $\mu_1 \sim \frac{M^2}{m_{pl,4}}$.  So long as 
$M_{np}/m_{pl,4} \ll M/\Lambda$ (where $M$ is the moduli mass and $\Lambda$ the
mass scale in \kskp), \survivalcond\ should be easily satisfied.  (Again, if 
we wished to consider the small effects
of these couplings, we should also consider couplings induced by supergravity effects.)

On the other hand, if we have a string theory model in which \modpotresc\ is determined
by (for example) some fluxes, then we can think of $\mu_0,\mu_1, c, \eps, \Lambda$ and so on 
in \survivalcond\ as depending on the value of these fluxes through 
the value $Z = z_0$ about which one expands to derive the effective theory
in \kskp,\Polonyisuper. If we replace Z with $\delta Z = Z - z_0$ in
\kskp,\Polonyisuper, then \survivalcond\ carves out a region of
the space of fluxes for which the effective field theory of 
$X, \delta Z$ describes SUSY-breaking dynamics.

Note that a problem potentially arises if the moduli mass is generated
by the same instantons as the coupling of the modulus to the SUSY-breaking dynamics.
In this case, $M \sim M_{np}^2/m_{pl,4} \sim \mu_1$, and \survivalcond\ is at best marginally
satisfied.  Furthermore, according to \fterms, the F-term for the modulus would be of the
same order as the F-terms for the "local model" of SUSY-breaking.  This is closer to 
moduli-domination scenarios as described in
\refs{\KaplunovskyRD\CasasQI\deCarlosDA\BrignoleDJ-\BrignoleFB}.  
In type II theories this could be avoided by burying the D-branes responsible
for SUSY-breaking down a warped throat, while ensuring that there are also D-instantons
supported away from the warped region which give the moduli a mass.
In addition, we could forbid the leading-order couplings with a discrete symmetry; terms which
are higher order in $\phi$ will be suppressed by additional powers of $m_{pl,4}$.

\newsec{Conclusions}

In the last two sections we have discussed cases where the SUSY-breaking
dynamics arises from fields distinct from the geometric moduli.
These might arise if the SUSY-breaking dynamics is "local": for example,
if it is localized on (intersecting) D-branes in type II string theory, or
on an "$E_8$ wall" in heterotic M-theory.

One can also imagine scenarios in which the moduli participate more directly.
One scenario for designing naturally small values of $m,\mu$ in the theory \orsuper\ 
is to couple the fields $X,A,B$ to non-abelian gauge fields \refs{\DineGM\DineDZ-\DineXT},
sometimes known as "retrofitting". 
Gaugino condensation introduces the small mass scales in such models. For example,
consider the case:
\eqn\retro{
	W = h X A^2 + \frac{a}{M_{\ast}} X W_{\alpha} W^{\alpha} + \frac{b}{M_{\ast}^2}
		AB \tilde{W}_{\alpha}\tilde{W}^{\alpha}
}
where $W,\tilde{W}$ are the chiral Fermi superfields for two $SU(2)$ gauge groups, and
$M_{\ast}$ is the cutoff scale of the theory, typically the Planck scale. Typically
$X,A,B$ are moduli themselves.  For example, if $W,\tilde{W}$ are open string gauge groups
in type IIB string theory, the gauge couplings will depend on the K\"ahler moduli of the theory.

Gaugino condensation will lead to the following low-energy superpotential:
\eqn\retrole{
	W = h X A^2 + \Lambda^{3} e^{\frac{c X}{M_{\ast}}} + \tilde{\Lambda}^3 
		e^{\frac{d AB}{M_{\ast}^2}}
}
where $c,d$ will depend on $a,b$.  Expanding the second and third terms to leading order,
we find the superpotential \orsuper\ plus higher-order corrections,\foot{In particular, there
will be a term $\eps X^2$; this will destabilize the SUSY-breaking pseudomoduli space
at tree level, apparently causing the system to roll to the supersymmetric point; however,
the Coleman-Weinberg potential will lead to a metastable SUSY-breaking vacuum near the origin,
so long as $\eps$ is sufficiently small.}
with $m = \frac{d \tilde{\Lambda}^3}{M_{\ast}^2}$ and $\mu = \frac{c \Lambda^3}{M_{\ast}}$.

Assume that $h \mu < m^2$ (which may take some doing). At this order, 
$A,B$ have masses of order $m$.  $X$ is stabilized by a one-loop mass
of order $m_{CW}^2 = \frac{h^4 \mu^2}{m^2} < \mu$ (if $h < 1$). Higher order 
corrections typically render these vacua metastable 
\refs{\DineGM\DineDZ-\DineXT,\IntriligatorDD,\IntriligatorCP}. 
Furthermore, the F-term for $X$ is nonvanishing
and is the order parameter for SUSY breaking.  If $X$ is a geometric
modulus, then this fits within the class of examples discussed in 
\refs{\KaplunovskyRD\CasasQI\deCarlosDA\BrignoleDJ-\BrignoleFB}, in which the F-terms
for the moduli provide the dominant contributions to the soft SUSY-breaking terms
(we are not addressing here the flavor problems which typically arise in these scenarios 
-- \cf\ \refs{\LouisHT} -- as well as in gravity mediation).
Indeed, much of the literature on gaugino condensation in string theory
makes use of this kind of scenario.

\vskip .3cm
\centerline{\bf Acknowledgements}

\vskip .2cm

I would like to thank Michael Dine, Shamit Kachru, John McGreevy, Erich Poppitz,
and Howard Schnitzer for discussions and for reading over a draft of this manuscript.
I would also like to thank Micha Berkooz, Oliver DeWolfe, 
Louis Leblond, Martin Schmaltz, Amit Sever, and
Eva Silverstein for helpful discussions.  Much of this work was completed at the
Aspen Center for Physics  during the summer 2008 workshop on "Supersymmetry breaking and
its mediation in field theory and string theory."  I would like to thank the
ACP staff and the organizers of the workshop for creating a stimulating
and productive working environment.
This work was supported by DOE Grant  
No.~DE-FG02-92ER40706, and by a 
DOE Outstanding Junior Investigator award.

\listrefs
\end